\documentstyle[12pt,aaspp4]{article} 
\begin{document}
\newcommand{\lya}{Lyman~$\alpha$}
\newcommand{\lyb}{Lyman~$\beta$}
\newcommand{\za}{$z_{\rm abs}$}
\newcommand{\ze}{$z_{\rm em}$}
\newcommand{\cmtwo}{cm$^{-2}$}
\newcommand{\nhi}{$N$(H$^0$)}
\newcommand{\nzn}{$N$(Zn$^+$)}
\newcommand{\ncr}{$N$(Cr$^+$)}
\newcommand{\degpoint}{\mbox{$^\circ\mskip-7.0mu.\,$}}
\newcommand{\halpha}{\mbox{H$\alpha$}}
\newcommand{\hbeta}{\mbox{H$\beta$}}
\newcommand{\hgamma}{\mbox{H$\gamma$}}
\newcommand{\kms}{\,km~s$^{-1}$}      
\newcommand{\minpoint}{\mbox{$'\mskip-4.7mu.\mskip0.8mu$}}
\newcommand{\mv}{\mbox{$m_{_V}$}}
\newcommand{\Mv}{\mbox{$M_{_V}$}}
\newcommand{\peryr}{\mbox{$\>\rm yr^{-1}$}}
\newcommand{\secpoint}{\mbox{$''\mskip-7.6mu.\,$}}
\newcommand{\sqdeg}{\mbox{${\rm deg}^2$}}
\newcommand{\squig}{\sim\!\!}
\newcommand{\subsun}{\mbox{$_{\twelvesy\odot}$}}
\newcommand{\et}{et al.~}
\newcommand{\cf}{c.f.~}
\newcommand{\eg}{e.g.~}

\def\ltsima{$\; \buildrel < \over \sim \;$}
\def\simlt{\lower.5ex\hbox{\ltsima}}
\def\gtsima{$\; \buildrel > \over \sim \;$}
\def\simgt{\lower.5ex\hbox{\gtsima}}
\def\arcs{$''~$}
\def\arcm{$'~$}

\title {REDSHIFT CLUSTERING IN THE HUBBLE DEEP FIELD\altaffilmark{1}}
\author{\sc Judith G. Cohen\altaffilmark{2},
Lennox L. Cowie\altaffilmark{3},
David W. Hogg\altaffilmark{4},
Antoinette Songaila\altaffilmark{3},
Roger Blandford\altaffilmark{4},
Esther M. Hu\altaffilmark{3} and Patrick Shopbell\altaffilmark{2}}

\altaffiltext{1}{Based in large part on observations obtained at the
W.M. Keck Observatory, which is operated jointly by the California 
Institute of Technology and the University of California}
\altaffiltext{2}{Palomar Observatory, Mail Stop 105-24,
California Institute of Technology}
\altaffiltext{3}{Institute for Astronomy, University of Hawaii, 2680
Woodlawn Drive, Honolulu, Hawaii 96822}
\altaffiltext{4}{Theoretical Astrophysics, California Institute of Technology,
Mail Stop 130-33, Pasadena, CA 91125}

\begin{abstract}
We present initial results from a redshift survey carried out with the
Low Resolution Imaging Spectrograph on the 10~m W. M. Keck Telescope
in the Hubble Deep Field.
In the redshift
distribution of the 140 
extragalactic objects in this sample we find 6
strong peaks, with velocity dispersions of ${\sim}400${\kms}. The
areal density of objects within a particular peak, 
while it may be non-uniform, does not
show evidence for strong central concentration.
These peaks have characteristics (velocity dispersions,
density enhancements, spacing, and spatial extent) similar to those 
seen in a comparable redshift survey in a different high galactic
latitude field (Cohen et al 1996), confirming that the structures 
are generic.  They are
probably the high redshift counterparts of huge galaxy structures
(``walls'') observed locally.
\end{abstract}


\section{INTRODUCTION}

The Hubble Deep Field (HDF hereafter; Williams et al 1996) has been surveyed to
extraordinary depths, with point source detection limits
around 29~mag in the $V$ and $I$ bands,
in an intensive
campaign by the Hubble Space Telescope in 1995 December.
The images represent the
deepest images ever taken in the optical and have already provided the
basis for studies of deep visual counts (Williams et al 1996), faint
object morphology (Abraham et al 1996), gravitational lensing (Hogg
et al 1996), and high-redshift objects (Steidel et al 1996; Clements
\& Couch 1996).  These studies represent only the beginning of a large
number of scientific projects possible with the HDF data.

In this paper we present the
first results of a ground based spectroscopic survey of galaxies
in the HDF with the Keck Telescope.  These observations were taken in
order to provide a database of object
redshifts for the use of the astronomical community and in order to
expand the faint object redshift surveys of Cowie et al (1996) and
Cohen et al (1996) to an additional field.

We assume an Einstein - de Sitter universe ($q_0 = 0.5$) with a
Hubble constant 100$h$ \kms Mpc$^{-1}$.

\section{REDSHIFT SAMPLE}

The HDF was selected on the basis of high galactic latitude, low
extinction, and various positional constraints described by Williams
et al (1996).  Redshifts were acquired with the Low Resolution Imaging
Spectrograph (Oke et al 1995) on the 10~m W. M. Keck Telescope over
two rectangular strips 2 x 7.3 arcmin$^2$ centered on the HST field in
1996 January, March and April.  One strip was aligned east-west
while the second was aligned at a position angle of 30$^{\circ}$ to
maximize the slit length that fell within the HDF itself, where the two
strips overlap.

The sample selection is different in each of 
the two strips.  The photometry and the definition of the sample
for spectroscopic work are described in Paper II of this series, Cowie
et al (1997). 
Plans exist to
complete the sample in a number of photometric bandpasses, but in view
of the great interest in the HDF and the many follow up studies
in progress, we present this data before the complete sample is available.

Table~1 presents the redshifts of 140 extragalactic objects, about half
of which are in the HDF itself and the remainder in the flanking fields.
The median redshift $z$ of the extragalactic objects in the present
sample is $z=0.53$.  Only three
are quasars or broad-lined AGNs.  12 Galactic stars were found as well. 
The radial velocity precision of our redshifts is unusually high for a
deep redshift survey.  We estimate that the uncertainty in $z$ for
those objects with redshifts considered secure and accurate
is $\approx 300$~\kms.  
Coordinates, crude ground based $R$ magnitudes in a 3 arc-sec diameter
intended for object identification only,
and redshifts are given in Table~1.

A more detailed account of the photometric and
spectroscopic properties of the entire sample including photometry
from $U$ through $K$  as well as a discussion of
incompleteness in the sample selection and redshift identification
is in preparation.
These incompletenesses ought not to affect the present work. 

\section{REDSHIFT DISTRIBUTION}

\subsection{Velocity Peaks}

The redshift histogram over the region 0.2 $< z < 0.9$ is 
is shown in Figure 1.  It shows clear evidence of clustering.  
Velocity peaks were identified by choosing bins of
variable width and centers so as to maximize their significance relative to occurring
by chance in a smoothed velocity distribution (smoothing width 20,000 \kms)
derived from the present
sample (c.f. Cohen et al 1996). 
Using this procedure we isolate 6 peaks significant
at better than 99.5~percent confidence (see Table 2). 
The fourth column in Table 2 gives a statistical significance parameter
$X_{max}$.  The fifth and sixth columns give the comoving transverse size
corresponding to 1 arc-min and the comoving radial distance corresponding to
${\Delta}z = 0.001$. 
The density in velocity
space within these peaks exceed the average density
by a factor that ranges from 4 to as high as 30
for the peak at $z_p = 0.321$.
40 percent of the total sample lies within these peaks.
Larger peaks including outliers are 
also highly significant.
The local velocity dispersions for these
peaks are strikingly small, ranging from 170 {\kms}~to 600 \kms. These are upper
bounds because they are comparable with our measurement errors.
They are also similar to the results obtained in a
high latitude field
for which we carried out a deep redshift survey with LRIS earlier 
(Cohen et al 1996).

By itself, this sample is too small to measure the two point correlation
function in velocity space. However, there is a $5\sigma$ excess correlation
in the 500--1000\kms ~interval with a correlation scale $V_0\sim600\pm200$\kms
(\cf Carlberg \et (1997), Le F\`evre \et 1996)
which can be converted into comoving distance along the line of sight using
the data in the sixth column of Table~2.  There is no evidence for
correlation with velocity differences in excess of 1000 \kms. No distinction between
low and high redshift is discernible.  There is no evidence for periodicity
in the peak redshifts (\cf Broadhurst et al 1990).


\subsection{Morphology Correlation}

If we make a simple morphological separation of the galaxies in the
redshift survey into spirals, ellipticals and
Peculiar/Mergers and use the HST images of the HDF 
and of
the flanking fields to classify these galaxies (c.f. van den Bergh et al 1996),
we find there is no indication of any
difference in population between the background field galaxies
and those in the redshift peaks.  In particular, the redshift
peaks do not contain a detectable excess of elliptical
galaxies.

\section{ANGULAR DISTRIBUTION}

The angular distribution of the entire sample and of the galaxies in
the two most populous velocity peaks is shown
in Figure~2.  The peculiar shape is caused by the use of
two LRIS strips with different position angles.  The outline of
the area covered is indicated by the solid lines, while the outline
of the area of the WFCII observations in the HDF
is indicated by the dashed lines.  The galaxies associated with the 6  
velocity peaks mostly exhibit 
a non-uniform distribution, though none show the strong central concentration
characteristic of clusters. The redshift sample must be completed before it 
is possible to make quantitative statements.



\subsection{Areal Density}

The areal density of galaxies brighter than $0.1L^*$ (as defined at $K$)
is computed for
redshift peaks in the 0 hour field (Cohen et al 1996) and for the
two largest peaks in the
HDF, where the $K$ photometry is not fully assembled yet.
Corrections have been applied for galaxies below the magnitude cutoff of the
survey assuming a flat luminosity function at the faint end.
To investigate a local analog to these structures, this is
repeated for the Local Group, for
the Virgo cluster (within a radius of 6$^{\circ}$ degrees from its
center)  using the survey of Kraan-Korteweg (1981)
and within the core of the Coma Cluster using data from 
Thompson \& Gregory (1980).  In these local structures,
the luminosity is determined at $B$
rather than at $K$.  The results are given in Table 3, and suggest that the best local analog
is the region of the Virgo cluster within 6$^{\circ}$ of its center,
but although the areal density is a reasonable match, the velocity
dispersion in the high redshift peaks is lower, often significantly
lower, than one sees in the central region of the Virgo cluster.

\section{DISCUSSION}

\subsection{Effects of sample definition decisions}

The conclusion of Cohen et al (1996), i.e., that a large
fraction of the galaxy population at redshifts to unity lie in low
velocity-dispersion structures, was based on a single field, but the
confirmation of strong redshift-space clustering in the HDF suggests
that the results are generic.  The clustering seen here is stronger
than that seen in other local and high-redshift surveys (Landy
et al 1996, LeF\`evre 1996, etc.)
The difference is attributed most importantly to the high sampling 
density in a small field.




\subsection{Structure Morphology}

At one level, these peaks may be no more than a manifestation
of the fact that galaxies are correlated in both configuration
and velocity space.  The connection between spatial and velocity correlation functions
is quite model-dependent (\eg Brainerd \et 1996).  Conversely, if
we can gain an empirical understanding of this relationship, it can
discriminate among cosmogonic models.  We briefly comment upon
some possibilities.

One explanation is 
that the velocity peaks represent
structures in velocity space and are not prominent in real space.
Such effects are sometimes seen in numerical simulations, \eg
Park \& Gott 1991, Bagla \& Padmanabhan 1994.
For example, they might be a ``backside infall'' into a large structure
where the Hubble expansion opposes the infall so as to give more or less
uniform recession velocity over a large interval of radial distance.
The generic kinematic difficulty with this explanation is that in order for 
features like this not to have many more descendants
in which the velocities have long ago crossed,
the characteristic lifetimes must be a significant fraction of the age of
the universe which, in turn, 
limits the mass density contrast to small 
values.  Given that half of the galaxies lie in these structures,
a large bias parameter must be invoked.

Alternatively, we may be observing structures that are spatially
compact and have the shapes of spheres, filaments, or walls.
We can
argue against these features being clusters on the following grounds:
(i) They do not exhibit central concentrations (\cf Sec.~4).
(ii) The velocity dispersions are too small, $200-600${\kms}~
as opposed to $600-1200$\kms.
(iii) The space density of rich clusters is too low;  the Palomar Deep
Cluster Survey (Postman et al 1996) finds only 7 clusters per square
degree out to $z\sim0.6$ with richness class $\ge1$. 
(iv) The redshift peaks do not show the excess of ellipticals
characteristic of rich clusters (Dressler 1980).

Small quasi-spherical groups are  
a possibility.  The mean free path is 
$\sim100 h^{-1}$ comoving Mpc.
The observed structures extend laterally over at least $\sim6$ arc-min
or $\sim2 h^{-1}$ Mpc, implying a space density
$\sim 3 {\rm{x}} 10^{-3} h^{3}$ Mpc$^{-3}$, $\sim$1/3 the density of $L^*$ 
galaxies.
Alternatively, we can associate the tentative velocity correlation
scale of $V_0\sim600$\kms~ with a radial extent of $\sim4 h^{-1}$~Mpc
and a lateral angular scale of $\sim12$ arc-minutes at $z \sim0.5$.

Filaments and walls have both been described in the theoretical
literature (\eg Bond et al 1996, Shandarin et al 1995).  Walls dominate
if there is excess power on large scales and are observed locally
(\eg in the Local Supercluster, deVaucouleurs 1975, and in local
redshift surveys, de Lapparant et al 1986, Landy et al 1996).
On this basis we speculate that the structures we are observing are
actually walls.

There are two obvious follow up investigations which can address this 
hypothesis.  The first is to perform similar redshift surveys 
in neighboring deep fields.  
If we assume that the wall normal
is inclined at an angle $\theta$ to the line of sight and that the 
constituent galaxies move with the Hubble flow in two dimensions,
then the variation of mean redshift with angular separation 
of the second survey $\Delta{\phi}$ and polar angle on the sky $\psi$ is 
$$
\Delta z=2[(1+z)^{3/2}-(1+z)]\Delta{\phi}\tan\theta\sin\psi
$$
For $z=0.5$, this is $\Delta z\sim2\times10^{-4}$ per arcminute
and in order to see redshift displacements in excess of the velocity 
dispersion, the additional surveys must be displaced by $\sim20'$.
With several lines of sight, it might be possible to test the 
above relation.

Secondly, wide field, multiband  photometric surveys to the depth of the 
redshift survey are clearly important to see if there are indeed morphological
and luminosity function differences between the galaxies within and outside
the velocity peaks.  Both investigations are underway.

\acknowledgements
We thank the Hubble Deep Field team, led by Bob Williams, for
planning, taking, reducing, and making public the HDF images.
We are grateful to George Djorgovski, Keith Matthews, Gerry
Neugebauer, Paddy Padmanabhan,
Mike Pahre, Tom Soifer and Jim Westphal for helpful conversations.
The entire Keck user community owes a huge
debt to Bev Oke, Jerry Nelson, Gerry Smith, and many other people who
have worked to make the Keck Telescope a reality.  We are grateful to
the W. M. Keck Foundation, and particularly its president, Howard
Keck, for the vision to fund the construction of the W. M. Keck
Observatory.  Support by NASA and the NSF is greatly appreciated.

\newpage

\begin{deluxetable}{llll|llll|llll}
\tablewidth{0pt}
\scriptsize
\tablecaption{
	Redshifts in the Hubble Deep Field
}
\tablehead{
   \colhead{RA}
 & \colhead{Dec}
 & \colhead{$R_{\rm ap}$}
 & \colhead{$z$}
 & \colhead{RA}
 & \colhead{Dec}
 & \colhead{$R_{\rm ap}$}
 & \colhead{$z$}
 & \colhead{RA}
 & \colhead{Dec}
 & \colhead{$R_{\rm ap}$}
 & \colhead{$z$}
\\
   \colhead{$(-12^{\rm h})$}
 & \colhead{$(-62^{\circ})$}
 & \colhead{$(3'')$}
 &
 & \colhead{$(-12^{\rm h})$}
 & \colhead{$(-62^{\circ})$}
 & \colhead{$(3'')$}
 &
 & \colhead{$(-12^{\rm h})$}
 & \colhead{$(-62^{\circ})$}
 & \colhead{$(3'')$}
 &
\\
   \colhead{m~s}
 & \colhead{$'~''$}
 & \colhead{mag}
 &
 & \colhead{m~s}
 & \colhead{$'~''$}
 & \colhead{mag}
 &
 & \colhead{m~s}
 & \colhead{$'~''$}
 & \colhead{mag}
 &
}
\startdata
36 21.4  & 12 27.1 & ---  & 0.398 &
36 22.0  & 12 37.7 & 21.7 & 0.630 &
36 22.2  & 12 41.9 & 20.8 & 0.498 \\
36 22.7  & 13 00.2 & 20.0 & 0.472 &
36 22.9  & 13 46.9 & 20.4 & 0.485 &
36 24.9  & 13 01.0 & 20.3 & 0.518 \\
36 26.5  & 12 52.6 & 20.6 & 0.557 &
36 27.7  & 12 41.3 & 20.8 & 0.518 &
36 28.1  & 12 38.0 & 21.1 & 0.5185 \\
36 29.8  & 14 03.8 & 21.4 & 0.793 &
36 29.9  & 12 25.0 & 22.6 & 0.410 &
36 30.2  & 12 08.8 & 20.6 & 0.456 \\
36 31.0  & 12 36.9 & 21.3 & 0.456 &
36 31.7  & 12 41.1 & 21.3 & 0.528 &
36 32.6  & 12 44.1 & 21.3 & 0.562 \\
36 33.4  & 13 20.3 & 21.1 & 0.843 &
36 33.04 & 11 35.0 & 19.4 & 0.080 &
36 33.6  & 11 56.8 & 21.8 & 0.458 \\
36 34.4  & 12 41.5 & 22.3 & 1.219 &
36 34.8  & 12 24.5 & 19.5 & 0.562 &
36 36.1  & 13 20.3 & 22.1 & 0.680 \\
36 36.3  & 13 41.2 & 21.4 & 0.556 &
36 36.78 & 11 36.1 & 19.4 & 0.078 &
36 37.2  & 12 53.1 & 20.8 & 0.485 \\
36 37.4  & 12 41.0 & 20.5 & 0.458 &
36 37.6  & 11 49.5 & 22.1 & 0.838 &
36 38.89 & 12 20.7 & 22.9 & 0.609 \\
36 39.8  & 12 07.5 & 21.8 & 1.015 &
36 40.80 & 12 04.4 & 23.7 & 1.010 &
36 41.56 & 11 33.1 & 20.5 & 0.089 \\
36 41.85 & 12 06.3 & 21.9 & 0.432 &
36 42.85 & 12 17.6 & 21.3 & 0.454 &
36 43.07 & 12 43.2 & 23.0 & 0.847 \\
36 43.55 & 12 19.4 & 23.4 & 0.752 &
36 43.69 & 13 57.7 & 21.6 & 0.201 &
36 43.71 & 11 44.0 & 22.3 & 0.765 \\
36 43.88 & 12 51.2 & 21.8 & 0.557 &
36 44.09 & 12 48.9 & 22.0 & 0.555 &
36 44.11 & 12 41.3 & 24.2 & 0.873 \\
36 44.28 & 11 34.3 & 23.2 & 1.013 &
36 44.59 & 12 28.8 & 24.2 & 2.268 &
36 45.32 & 12 14.5 & 21.4 & 0 \\
36 45.86 & 12 02.4 & 24.6 & 0.679 &
36 46.10 & 11 42.9 & 22.6 & 1.016 &
36 46.25 & 14 05.6 & 22.6 & 0.960 \\
36 46.44 & 11 52.3 & 22.9 & 0.5035 &
36 46.45 & 14 08.6 & 23.1 & 0.130 &
36 46.68 & 12 38.1 & 23.0 & 0.320 \\
36 46.78 & 11 45.9 & 23.1 & 1.059 &
36 47.21 & 12 31.8 & 23.4 & 0.421 &
36 47.99 & 13 10.1 & 21.5 & 0.475 \\
36 48.5  & 13 29.2 & 23.9 & 0.958 &
36 48.51 & 11 42.3 & 23.2 & 0.962 &
36 49.29 & 13 12.3 & 22.7 & 0.478 \\
36 49.34 & 13 47.9 & 19.0 & 0.089 &
36 49.42 & 14 07.8 & 22.8 & 0.752 &
36 49.55 & 12 58.8 & 22.6 & 0.475 \\
36 49.64 & 13 14.2 & 22.4 & 0.475 &
36 50.15 & 12 40.8 & 21.4 & 0.474 &
36 50.18 & 12 46.9 & 22.8 & 0.680 \\
36 50.63 & 10 59.9 & 21.9 & 0.474 &
36 50.73 & 12 56.9 & 23.1 & 0.320 &
36 51.0  & 13 21.6 & 20.8 & 0.199 \\
36 51.02 & 10 32.2 & 21.2 & 0.410 &
36 51.35 & 13 01.6 & 22.2 & 0.089 &
36 51.61 & 12 21.3 & 22.3 & 0.299 \\
36 51.69 & 13 54.8 & 22.0 & 0.557 &
36 52.03 & 14 58.3 & 22.4 & 0.358 &
36 52.39 & 10 36.9 & 22.2 & 0.321 \\
36 52.59 & 12 21.0 & 24.0 & 0.401 &
36 52.68 & 13 55.7 & 22.7 & 1.355 &
36 52.71 & 14 32.9 & 21.2 & 0 \\
36 52.83 & 14 54.7 & 22.7 & 0.463 &
36 52.85 & 14 45.1 & 20.1 & 0.322 &
36 53.33 & 12 35.2 & 23.4 & 0.560 \\
36 53.54 & 15 26.0 & 18.7 & 0 &
36 53.57 & 13 09.4 & 22.1 & 0 &
36 53.77 & 12 55.0 & 22.0 & 0.642 \\
36 54.28 & 14 35.1 & 22.8 & 0.577 &
36 54.65 & 13 29.1 & 20.0 & 0 &
36 55.44 & 13 54.5 & 22.4 & 1.148 \\
36 55.45 & 12 46.4 & 23.1 & 0.790 &
36 55.50 & 14 00.9 & 23.9 & 0.559 &
36 56.26 & 12 42.4 & 19.9 & 0 \\
36 56.33 & 12 10.4 & 23.7 & 0.321 &
36 56.56 & 12 46.8 & 21.7 & 0.5185 &
36 57.14 & 12 27.1 & 23.4 & 0.561 \\
36 57.22 & 13 00.8 & 22.3 & 0.474 &
36 57.64 & 13 16.5 & 23.8 & 0.952 &
36 57.98 & 13 01.6 & 23.0 & 0.320 \\
36 58.22 & 12 15.2 & 22.9 & 1.020 &
36 58.29 & 15 49.4 & 21.7 & 0.457 &
36 58.56 & 12 23.0 & 24.2 & 0.682 \\
36 58.64 & 14 39.1 & 23.3 & 0.512 &
36 58.66 & 12 53.2 & 22.2 & 0.321 &
36 58.74 & 14 35.6 & 21.9 & 0.678 \\
36 58.76 & 16 38.9 & 20.0 & 0.299 &
36 59.43 & 12 22.7 & 24.5 & 0.472 &
36 59.79 & 14 50.6 & 22.5 & 0.761 \\
37 00.41 & 14 06.7 & 21.5 & 0.423 &
37 00.47 & 12 35.9 & 24.5 & 0.562 &
37 01.8  & 13 23.8 & 20.7 & 0.408 \\
37 01.81 & 15 10.9 & 22.9 & 0.938 &
37 02.3  & 13 43.0 & 21.3 & 0.559 &
37 02.5  & 13 48.3 & 22.7 & 0.513 \\
37 02.5  & 14 02.7 & 22.1 & 1.243 &
37 02.70 & 15 44.8 & 20.8 & 0.514 &
37 02.81 & 14 24.4 & 21.5 & 0.512 \\
37 03.21 & 16 46.9 & 23.0 & 0.744 &
37 03.6  & 13 54.3 & 21.7 & 0.745 &
37 03.82 & 14 42.0 & 22.3 & 0.475 \\
37 03.91 & 15 23.8 & 22.6 & 0.377 &
37 04.17 & 16 25.3 & 22.8 & 0.474 &
37 04.52 & 16 52.2 & 21.1 & 0.377 \\
37 04.56 & 14 30.0 & 22.0 & 0.561 &
37 04.73 & 14 55.8 & 21.2 & 0 &
37 04.91 & 15 47.4 & 23.4 & 0.533 \\
37 05.0  & 12 11.2 & 22.5 & 0.386 &
37 05.66 & 15 25.7 & 22.7 & 0.503 &
37 06.0  & 13 33.9 & 21.6 & 0.753 \\
37 06.81 & 14 30.3 & 21.2 & 0 &
37 07.0  & 12 14.7 & 21.4 & 0.655 &
37 07.0  & 11 58.5 & 22.4 & 0.593 \\
37 07.73 & 16 06.1 & 22.8 & 0.936 &
37 08.01 & 16 31.7 & 22.7 & 0 &
37 08.04 & 16 59.6 & 21.5 & 0.458 \\
37 08.1  & 12 53.2 & 21.9 & 0.838 &
37 08.1  & 13 21.6 & 22.7 & 0.785 &
37 08.20 & 14 54.8 & 22.8 & 0.565 \\
37 08.25 & 15 15.3 & 22.5 & 0.839 &
37 08.53 & 15 02.2 & 22.7 & 0.570 &
37 08.60 & 16 12.4 & 21.3 & 0 \\
37 08.8  & 12 02.8 & 22.6 & 0.855 &
37 09.46 & 14 24.3 & 22.0 & 0.476 &
37 09.79 & 15 25.0 & 20.0 & 0.597 \\
37 10.1  & 13 20.5 & 21.7 & 0.320 &
37 11.85 & 16 59.7 & 23.5 & 1.142 &
37 12.4  & 13 58.2 & 22.6 & 0.848 \\
37 12.58 & 15 43.4 & 22.3 & 0.533 &
37 13.0  & 13 57.2 & 22.0 & 1.016 &
37 13.59 & 15 12.0 & 22.1 & 0.524 \\
37 14.8  & 13 35.4 & 22.5 & 0.897 &
37 16.1  & 13 54.2 & 21.5 & 0.476 &
37 16.32 & 16 30.4 & 23.4 & 0 \\
37 16.4  & 13 11.2 & 21.9 & 0.898 &
37 17.0  & 13 57.4 & 20.7 & 0.336 &
37 16.52 & 16 44.7 & 22.7 & 0.557 \\
37 18.28 & 15 54.1 & 21.6 & 0.476 &
37 18.3  & 13 48.6 & 22.1 & 0.480 &
37 18.4  & 13 22.5 & 20.6 & 0.4755 \\
37 18.60 & 16 05.0 & 22.5 & 0.558 &
37 22.25 & 16 13.1 & 22.6 & 0
\enddata
\end{deluxetable}

\newpage
\begin{deluxetable}{crcrrr}
\tablewidth{0pc}
\scriptsize
\tablecaption{Redshift Peaks in the Hubble Deep Field}
\tablehead{
\colhead{$z_p$} & \colhead{$N$\tablenotemark{a}} & 
\colhead{$\sigma_v(N)$\tablenotemark{b}} &
\colhead{$X_{max}$\tablenotemark{c}} 
& \colhead{$d_{\bot}$ ($\Delta(\theta)$=1')} & 
\colhead{$d_{\Vert}$ ($\Delta(z)$=0.001)} \nl
\colhead{~} & \colhead{~}  & \colhead{(\kms)} &   ~
& \colhead{($h^{-1}$ Mpc)\tablenotemark{d}} &
\colhead{($h^{-1}$ Mpc)\tablenotemark{d}} }
\startdata
0.321 &  8 &  170 &  22  & 0.22 & 2.0 \nl
0.457 &  7 &  310 &  10  & 0.31 & 1.7 \nl
0.475 & 15 &  315 &  21  & 0.31 & 1.6 \nl
0.516 &  8 &  595 &   8  & 0.33 & 1.6 \nl
0.559 & 14 &  420 &  21  & 0.34 & 1.5 \nl
0.680 &  5 &  265 &   8  & 0.40 & 1.4 \nl
\enddata
\tablenotetext{a}{Number of galaxies within the peak as determined
by statistical tests.}
\tablenotetext{b}{No correction for instrumental or measurement
errors has been applied.}
\tablenotetext{c}{Statistical parameter for estimating the significance
of each peak, see Cohen et al 1996.}
\tablenotetext{d} {Comoving distances.}
%
%
%
\end{deluxetable}

\clearpage
\newpage
\begin{deluxetable}{crrrc}
\tablewidth{0pc}
\scriptsize
\tablecaption{Areal Density of Peaks in The Caltech 0 hour Field and in
Local Structures}
\tablehead{
\colhead{$z_p$} & \colhead{N$_{\rm obs}$} & 
\colhead{Comoving Area} & 
\colhead{n$_{\rm corr}$($L>0.1L^*$)\tablenotemark{g}}
& \colhead{$\sigma_v$(N-1)} \nl
\colhead{~} & \colhead{($L>0.1L^*$)} & \colhead{($h^{-2}$ Mpc$^2$)} 
&\colhead{($h^2$ Mpc$^{-2}$)} &\colhead{(\kms)}}
\startdata
0.392 &  3 & 1.03 &  3 & 465 \nl
0.429 & 14 & 1.19 & 13 & 615 \nl
0.581 & 23 & 1.86 & 19 & 410 \nl
0.675 &  8 & 2.30 &  7 & 405 \nl
0.766 &  7 & 2.72 &  7 & 670 \nl
0.475 (HDF)\tablenotemark{f} &  7 & 0.51 & 18 & 315 \nl
0.559 (HDF)\tablenotemark{f} &  7 & 0.64 & 17 & 420 \nl
Local Structures \nl
Local Group & 4 & 1.3 & 3\tablenotemark{a} & $<100$ \nl
Virgo\tablenotemark{b} & 122 & 8.5 & 14\tablenotemark{a} & 670
\tablenotemark{d} \nl
Coma\tablenotemark{c} & 248 & 2.0 & 125 & 1080 \tablenotemark{e} \nl
\enddata
\tablenotetext{a}{Independent of $h$.}
\tablenotetext{b}{Galaxies within a 6$^{\circ}$ radius of the cluster center.}
\tablenotetext{c}{Galaxies within the central region $1.2^{\circ}$ on a side}
\tablenotetext{d}{Bingelli, Sandage \& Tammann (1985)}
\tablenotetext{e}{Colless \& Dunn (1996) (square region $2.6^{\circ}$ 
on a side)}
\tablenotetext{f}{The area is that of the 3 WF CCDs.}
\tablenotetext{g}{Comoving areal density corrected for incompleteness
at the faint end.}
%
\end{deluxetable}
%
%
%

\clearpage 
\figcaption[/scr2/jlc/deep0hr/paper1/figure1.ps]
{The redshift histogram for the galaxies in the merged Caltech
and Hawaii survey of the HDF.
\label{fig1}}
\figcaption[/scr2/jlc/deep0hr/paper1/figure2.ps]
{The distribution of our sample of galaxies projected
onto the sky is shown. 
Galaxies in the two most populous redshift peaks are indicated.
\label{fig2}}

\clearpage
\begin{figure}
\plotone{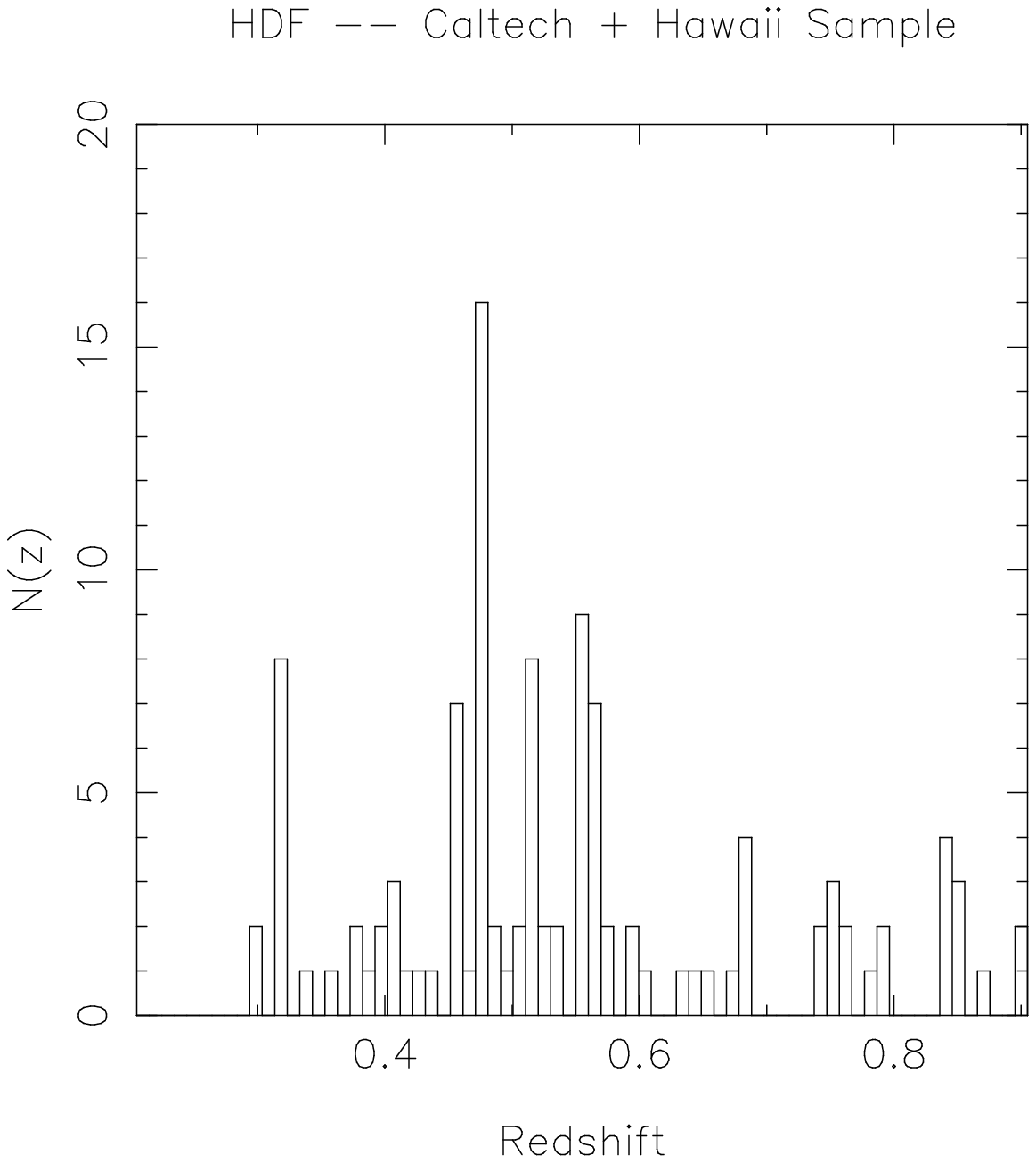}
\end{figure}
\clearpage
\epsscale{0.9}
\begin{figure}
\plotone{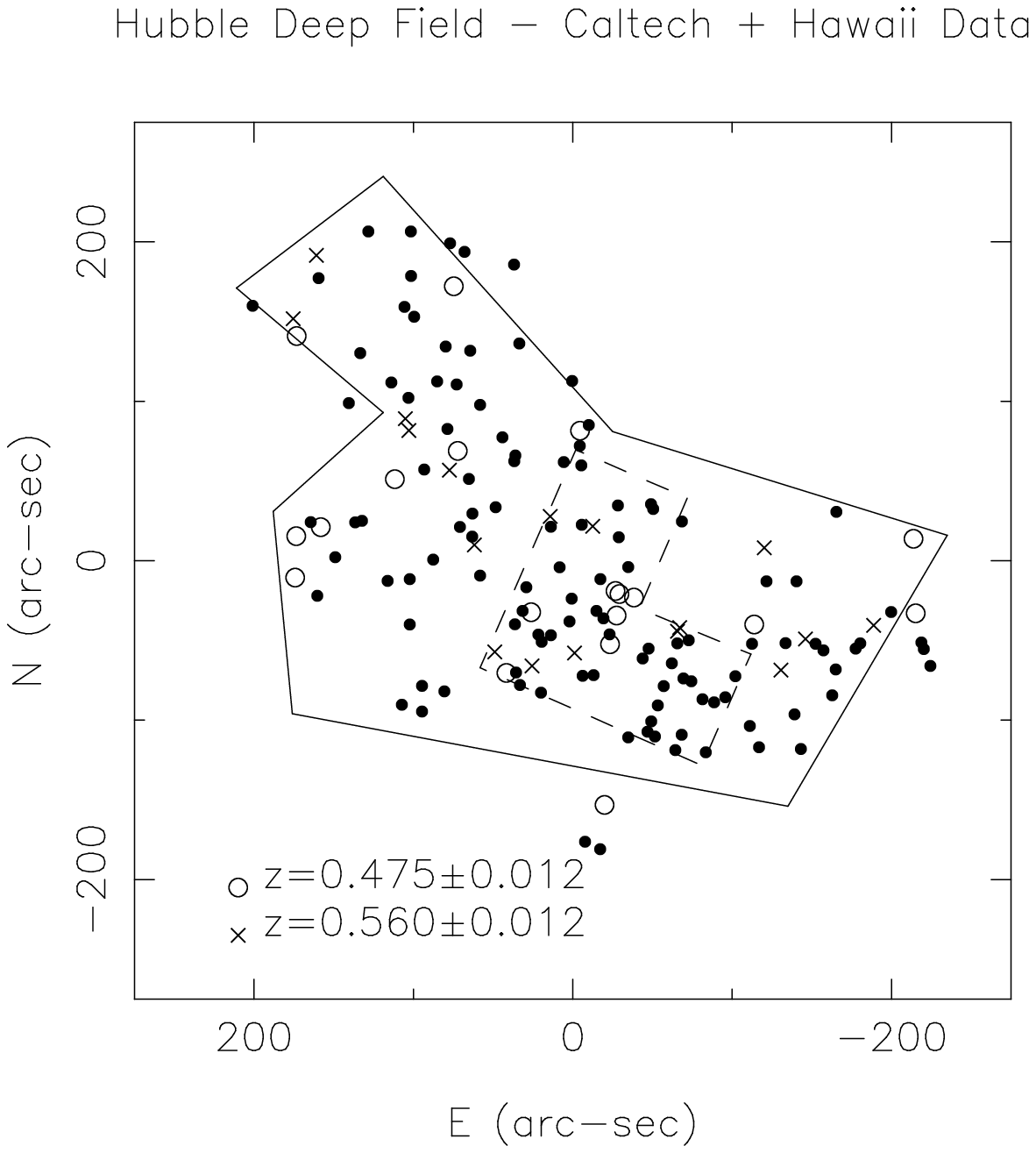}
\end{figure}
\end{document}